\documentclass[12pt]{article}
\usepackage{latexsym}
\usepackage{graphics}
\usepackage{epsfig}
\usepackage[mathscr]{eucal}
\usepackage{amscd}
\usepackage{amssymb}
\usepackage{amsmath}

 at10pt

\newcommand{\hs}{\Omega_{2D}}

\newcommand{\pd}{\partial}
\newcommand{\re}{\mathbb R}
\newcommand{\cV}{\mathcal V}

\newcommand{\tZ}{\widetilde Z}

\newcommand{\bM}{\mathbb M}

\newcommand{\id}{\mathrm 1}
\newtheorem{theorem}{Theorem}
\newtheorem{claim}{Claim}
\newcommand{\bproof}{\setlength{\parindent}{0mm}{\bf Proof{~~}}}
\newcommand{\eproof}{\hfill $\Box$\setlength{\parindent}{5mm}}
\title{A 2D model of Causal Set Quantum Gravity: \\ The emergence of
 the continuum.}

\author{Graham Brightwell${}^1$, Joe Henson${}^2$ and Sumati Surya${}^3$ \\
${}^1$London School of Economics, London, UK, \\
${}^2$Perimeter Insitute, Waterloo, Canada \\ 
\& University of Utrecht, Utrecht, Netherlands,  \\
${}^3$Raman Research Institute, Bangalore, India }

\begin{document}
\maketitle
\begin{abstract}
Non-perturbative theories of quantum gravity inevitably include
configurations that fail to resemble physically reasonable spacetimes
at large scales.  Often, these configurations are entropically
dominant and pose an obstacle to obtaining the desired
classical limit.  We examine this ``entropy problem'' in a
model of causal set quantum gravity corresponding to a discretisation
of 2D spacetimes.  Using results from the theory of partial orders we
show that, in the large volume or continuum limit, its partition
function is dominated by causal sets which approximate to a region of
2D Minkowski space. This model of causal set quantum gravity thus
overcomes the entropy problem and predicts the emergence of a
physically reasonable geometry.
\end{abstract}

In approaches to quantum gravity where the continuum is replaced by a
more primitive entity, manifoldlikeness is typically a feature of only
a small proportion of the configurations. In order to obtain the
correct continuum limit, this small set of configurations needs to be
dynamically favoured over the often overwhelming entropic contribution
from non-manifoldlike configurations.  It has been argued that some
form of this ``entropy problem'' is of critical importance in
dynamical triangulations, graph-based approaches  and in causal set
quantum gravity(CSQG) \cite{Entropy}.  This present work 
shows how the problem is overcome in a simplified 2D model of CSQG.

In CSQG, continuum spacetime arises as an approximation to a
fundamentally discrete structure, the causal set. Here, order and
number correspond to the continuum notions of causal order and
spacetime volume.  Despite being discrete, local Lorentz invariance in
the continuum approximation is restored by using a random lattice
\cite{bhs}.  These basic features of the theory led to an early
prediction for the cosmological constant, confirmed several years
later by observation \cite{lambda}. The construction of a model of
CSQG with physically realistic predictions is therefore of
considerable interest.
 
The fundamental entity that replaces spacetime in CSQG is a causal
set, or causet, $(C,\prec)$, which is a locally finite partially
ordered set. Namely, for any $x,y,z \in C$ (i) $x \nprec x$
(irreflexive)\footnote{This can be replaced by the condition that $x
\prec y $, $y \prec x \Rightarrow x=y$. For both choices one avoids
causal ``loops''.}, (ii) $x \prec y$, $y \prec z$ $\Rightarrow x \prec
z$ (transitive) and (iii) Cardinality($\mathrm{Future}(x) \cap
\mathrm{Past}(y)$) is finite (locally finite), where
$\mathrm{Future}(x)= \{z| x \prec z \} $, and $\mathrm{Past}(x)=\{ z |
z\prec x \}$. Local finiteness means that discreteness is taken to be
fundamental, and not simply a tool for regularisation.  The continuum
$(M,g)$ arises as an approximation of a causet $C$ if there exists a
``faithful embedding'' $\Phi:C \rightarrow (M,g)$ at density
$V_c^{-1}$, where $V_c$ is the discreteness scale.  This means that
(a) the distribution of $\Phi(C) \subset M$ is indistinguishable from
that obtained via a Poisson sprinkling into $(M,g)$, i.e., a random
discrete set $S$ such that, for any region $R$ of volume $V$, the
number of points of $S$ in $R$ is a Poisson random variable with mean
$V/V_c$, and (b) the order relation $\prec$ in $C$ is in 1-1
correspondence with the induced causal order on $\Phi(C)$, i.e., $x
\prec y \Leftrightarrow$ $\Phi(x)$ is to the causal past of $\Phi(y)$
\cite{causet,luca}.

While a quantum version of a classical sequential growth dynamics for
causets may eventually provide a more natural framework for
quantisation \cite{DR}, it is useful to consider the standard
path-integral paradigm.  As in other discrete approaches
\cite{statesum}, the path integral is replaced by a sum, which in CSQG
is over causets, with an appropriate ``causet action'' providing the
measure. The Regge action is an obvious choice for discrete theories
of quantum gravity based on triangulations of spacetimes \cite{regge}.
However, in CSQG, because of the intrinsic non-locality of causets, an
action defined as a sum of strictly local quantities is likely to
fail. The construction of a causet action is deeply intertwined with
the question of locality and the associated problems in constructing a
consistent quantum dynamics \cite{Joe,Dalem}.  While prescriptions
for a localised D'alembertian \cite{Dalem} may eventually lead to an
approximately local action for causets, it is a worthwhile exercise to
sidestep this question by considering simplified models.

One possible approach is to make a precise definition of the class of
``manifoldlike'' causets, and restrict the history space to this
class. Such causets have a natural corresponding continuum action
which can be used to define the partition function. While
manifoldlikeness is a trivial prediction of such a model, it may
nevertheless display features that yield interesting insights into
CSQG. Without further restrictions, however, such a model is not
obviously tractable. One such restriction is by spacetime dimension,
the simplest choice being dimension 2.

The specific model of 2D CSQG we present here is constructed via a
restriction to the class of so-called 2D orders.  This class contains
not only all causets that approximate to conformally flat 2D spacetime
intervals, but also some that are non-manifoldlike. Moreover, all
causets in this class share a certain topological triviality.  This
allows us to meaningfully address the entropy problem and the question
of manifoldlikeness within the model. We find that the entropy problem
is tamed in an unexpected way, and that it is possible to characterise
its physical consequences with results that may be surprising.

Such a model can be regarded as a restriction of the full theory of
CSQG, analogous to mini and midi-superspace models in canonical
approaches to quantum gravity -- the hope would be to gain insights
into the full theory by understanding details of the simplified
model. 
Indeed, our model is more fully dynamical than such reduced
models, because rather than freezing local degrees of freedom, one is
simply restricting to a  class of causal sets that are naturally
associated with discretisations of 2D spacetime, with a fixed topology. 
Causal set theory does not in principle assume a fixed spacetime
dimension, and hence our 2D model is indeed a restriction of the full
theory. However, this simply brings it on par with the starting point
of other routes to quantisation which must assume a fixed spacetime
dimension. While a restriction to topology is routinely adopted in
other approaches to quantum gravity, the hope is that our model can
ultimately be generalised to include a sum over all 2D topologies.

Although our work does not lead directly to a 4D theory, it is
an example of how the continuum can be recovered from a quantum causet
model and hence may prompt more optimism on the general approach
presented above.  Moreover, it is an explicit demonstration that the
causet approach is rich enough to allow formulations with physically
sensible outcomes, without the addition of extra variables
\cite{sfcs}.

We  consider a causet ``discretisation'' of the set of 2D conformally flat
spacetime intervals $(I,g)$. Using a fiducial flat metric, $\eta_{ab}
dx^a dx^b = -du dv$ in light cone coordinates $(u,v)$, these
geometries are represented by diffeomorphism classes of the metrics
\begin{equation}\label{flatclass}
g_{ab} dx^a dx^b = - \Omega^2(u,v)\,du dv,  
\end{equation}
with $\Omega^2(u,v)$ the conformal factor.  Quantisation of this set
of spacetimes on $I$ can be thought of as a  Lorentzian analog of Euclidean 2D quantum
gravity on a disc \cite{fdavid}. As a topological space, $I$ is simply
homeomorphic to a disc, with the boundary condition that there exists
an interval $I_0$ of ${}^2\bM$ and a bijection $\Psi: \pd I
\rightarrow \pd I_0$.  In 2D the conformal factor encodes all
geometric degrees of freedom so that all Lorentzian metrics on this
manifold have the form (\ref{flatclass}).

We will also adopt the unimodular modification of gravity, in which
spacetime volume plays the role of a time parameter
\cite{unimod}. Fixing the time coordinate is thus given a covariant
meaning, corresponding to the volume constraint 
\begin{equation} 
\cV = \int
\Omega(u,v) \, \, dudv = \mathrm{constant}.  
\end{equation} 
This constraint places restrictions on the map $\Psi$; starting with
an ``initial'' event $p_0 = (u_0, v_0)$ in the fiducial metric
$\eta_{ab}$, the ``final'' event $p_f=(u_f, v_f)$ of the interval $I_0
= [(u_f, v_f), (u_0,v_0) ]$ in $(\re^2,\eta_{ab})$ is determined (upto
boosts) by the condition $\int_{u_0}^{u_f} \int_{v_0}^{v_f}
\Omega(u,v) \, du dv = \cV.$

The Einstein action on an interval includes a term on the null
boundary. In order to simplify the action, we take 
the interval $I$ to be enlarged ever so slightly, to a region $I' \supset I$ with
spacelike boundary components. Because of the nature of the manifold
approximation, a causet discretisation is insensitive to modifications
of the boundary on scales much smaller than the discretisation scale,
i.e. if the volume of the region $I'-I$ is $\ll V_c$. Since the
boundary of $I'$ is piecewise spacelike, the Einstein action takes the
form  
\begin{equation} 
S  =   \frac{1}{16 \pi G} \int_{I'} R dV -  \frac{1}{8 \pi G} \int_{\pd
  I'} k dS -  \sum_{j} \frac{1}{8 \pi G} \theta_j - \frac{1}{8 \pi G}
  \Lambda V_{I'}, \label{contaction} 
\end{equation}  
where $\theta_j$ are the four boost parameters corresponding to the
four ``joints'' in $\pd I'$ \cite{hayward,GB}.  The Lorentzian  
Gauss Bonnet theorem \cite{BN,Law} then simplifies the action to 
\begin{equation} 
S = \frac{1}{16 \pi G} ( 2 \pi i  
  -  2   \Lambda  V_{I'}),
 \end{equation}  
which is a constant over the entire class of spacetimes under
consideration.   

The first step in constructing the causet discretisation of this
continuum theory is to characterise the class of causets of finite
cardinality which embed faithfully into conformally flat 2D intervals.
While such a characterisation appears daunting in general, for our
model these causets lie in the set of ``2D orders'', a well-studied
class of partially ordered sets.

\begin{figure}[ht]
\centering \resizebox{3.5in}{!}{\includegraphics{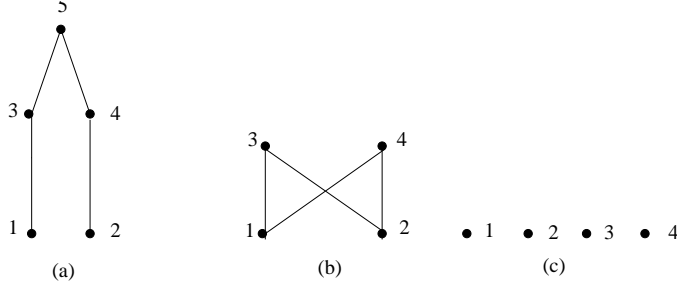}}
\vspace{0.5cm}
\caption{{\small Examples of labelled 2D orders, obtained from the
    intersections of the following linear orders: (a)
    $L=(e_1,e_3,e_2, e_4,e_5) $ and $M=(e_2,e_4,e_1,e_3,e_5)$
    (b) $L=(e_1,e_2,e_3, e_4) $ and $M=(e_2,e_1,e_4,e_3)$ and
    (c) $L=(e_1,e_2,e_3, e_4) $ and $M=(e_4,e_3,e_2,e_1)$.}}\label{twodorders}
\end{figure}

To define this class, some nomenclature is necessary.  Consider a set
of elements $S=\{e_1,...e_n\}$ and a partial order $\prec$ on this
set.  A causet $X$ on the underlying set $S$ is a linear order if and
only if, for all $i,j$, $e_i \prec e_j$ or $e_j \prec e_i$ in $X$.
We use the notation $Q=(e_{\pi(1)},e_{\pi(2)}\ldots e_{\pi(i)},
\ldots e_{\pi(n)})$ to denote a linear order on $S$, where $\pi$ is a
permutation on $n$ elements, so that $e_{\pi(i)} \prec e_{\pi(i+1)}$
for all $i$.  A linearly ordered subset of a causet is known as a
chain.  Similarly, a totally unordered causet is one such that $e_i
\nprec e_j$ for all $i,j$, and a totally unordered subset is known as
an antichain.  For causets $Q_1, Q_2, \dots, Q_k$ on the same set $S$,
the intersection $P= \bigcap_{i=1}^k Q_i$ is defined by setting $e_i
\prec e_j$ in $P$ if and only if $e_i \prec e_j$ in all of the $Q_i$.
For example, if $Q_1=(e_1,e_2)$ and $Q_2=(e_2,e_1) $, then since $e_1
\prec e_2 $ in $Q_1$ and $e_2 \prec e_1 $ in $Q_2$, they are unrelated
the intersection $Q_1\cap Q_2$ which is therefore a two element
antichain. The ``dimension'' of a causet $P$ is the minimum $k$ such
that $P$ can be written as the intersection of $k$ linear orders.  Our
main interest is in ``2-dimensional'' or 2D orders: ones that can be
written as the intersection of two linear orders, but are not
themselves linear orders\footnote{Note that causet dimension in this
context is not apriori related to the spacetime dimension.} (see Fig
\ref{twodorders}).

In the rest of this paper we will say that a causet $C$
``corresponds'' to a given spacetime $(M,g)$ if there exists a
faithful embedding $\Phi:C \rightarrow (M,g)$, defined precisely by
Bombelli as follows \cite{luca}. Let $\Phi:C \rightarrow M$ be an
embedding of a causet $C$ of cardinality $ V/V_c$ into a spacetime of
finite volume $V$. Consider sampling intervals of volume $V_c< V_0 <
V$. Then 
\begin{equation} 
P_{V_0}(n) \equiv  \frac{1}{n!} e^{-\frac{V_0}{V_c}}
\Biggl(\frac{V_0}{V_c}\Biggr)^{n} 
\end{equation} 
is the probability of finding $n < N $ elements of $\Phi(C)$ in a
region of volume $V_0$ for a Poisson embedding.  Define the indicator
function $F_n= \int \chi_n(I) dI/\int dI$, for the embedded causet
$\Phi(C)$, where $\chi_n(I)=1$, or $0$ depending on whether the
interval $I$ (of volume $V_0$) has $n$ points in it or not, and the
integral is over all possible intervals $I$ of volume $V_0$ in
$(M,g)$. Then, if $|F_n - P_{V_0}| < \delta$, $\Phi$ will be said to
be a $\delta$-faithful embedding with respect to $V_0$. We will
henceforth use the phrase ``faithfully embeddable'' to imply in the
$(\delta, V_0)$ sense. Specifically, we will require that $V_c \ll V_0
\ll V$ and $0< \delta \ll 1 $. For suitable choices of $\delta$ and
$V_0$, a causet generated by a Poisson sprinkling into $M$ 
 with density $V_c^{-1}$ will be, with high probability,
faithfully embedded in $M$.  On the other hand, regular discrete
lattices tend not to be faithfully embedded: the regular structure
leaves large intervals void of points.

To see that 2D orders are appropriate for our purposes, consider a
conformally flat 2D spacetime $(M,g)$.  The causal order $\preceq$
between events $p$ and $q$ in such a spacetime can be encoded in the statement:
 \begin{equation} \label{order}
 (u_1, v_1) \preceq (u_2,v_2) \quad \Leftrightarrow \quad u_1 \leq u_2 \quad
  {\mathrm{and}}\quad v_1 \leq v_2, 
 \end{equation}
where $(u_{1,2},v_{1,2})$ are light cone coordinates of $p$ and $q$,
respectively.  For conformally flat spacetimes any choice of light
cone coordinates is such that the ordering on each co-ordinate $u$ or
$v$ is a linear order.  This means exactly that a finite causet can be
embedded in $(M,g)$ if and only if it is the intersection of the two
co-ordinate linear orders, i.e., if and only if its ``dimension'' is
at most~2 \cite{trotter}.

Although every 2D order can be embedded into a conformally flat 2D
spacetime, not all of them can be \textit{faithfully} embedded. For
example, the intersection of the linear orders $L=(e_1, e_2, e_3, e_4,
\ldots, e_N) $ and $M=(e_2, e_1, e_3, e_4, \ldots, e_N)$ has an
antichain $ \{ e_1,e_2 \}$, while all other $e_i$ are to the future of
both $e_1$ and $e_2$, and linearly ordered. Thus $L\cap M$ is
\textit{almost} a chain, except for the past-most two elements $\{e_1,
e_2 \}$, and so clearly does not faithfully embed into a 2D spacetime,
at least for $N$ sufficiently large. Thus, in this sense, not every 2D
order corresponds to a 2D spacetime.

\begin{figure}[ht]
\centering \resizebox{3.5in}{!}{\includegraphics{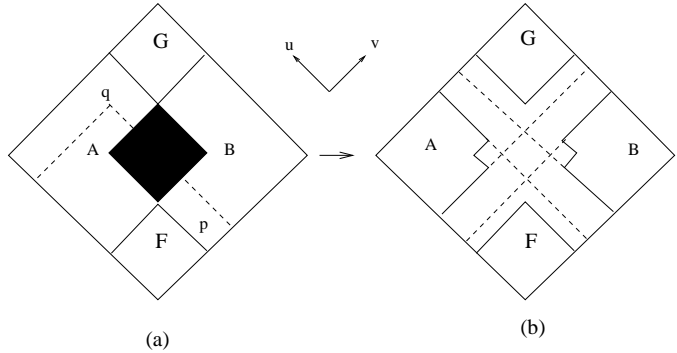}}
\vspace{0.5cm}
\caption{{\small (a) If $\Phi: C \rightarrow (I-R,\eta)$ is faithful,
    $\Phi(C)$ uniformly populates the regions $A,B,F,G$. (b) A
    suitable change in the density of the embedding pushes the
    elements of $C$ in regions $A,B,F,G$ of (a) into a portion of the
    spacetime $(I,\eta)$ without changing the order-causality
    correspondence. By choosing a conformal factor $\Omega(u,v)$ which
    is approximately one in the regions $A,B,F,G$ and vanishingly small
    elsewhere, $\Phi(C)$ can be equivalently thought of as a faithful
    embeddedding into $(I,\Omega^2(u,v) \eta)$.}}\label{difftgy}
\end{figure}

But can a 2D order faithfully embed into a spacetime of a different
topology than the interval?  Consider, for example, a flat 2D interval
$(I,\eta)$ with a large region $R$ cut out of it (see Fig
\ref{difftgy} (a)). $(I-R,\eta)$ is not a causally convex subset of
$(I,\eta)$ and hence its intrinsic causal order differs from that of
$(I,\eta)$. In particular, it contains pairs of events $p=(u_1,v_1),
q=(u_2,v_2)$ such that $p \prec q$ in $(I,\eta)$, but $ p \nprec q$ in
$(I-R,\eta)$, so that $u_1<u_2$, $v_1<v_2$ does not imply that
$(u_1,v_1) \prec (u_2,v_2)$.  This means that a causal set $C$ that
faithfully embeds, via some $\Phi$, into $(I-R,\eta) $ cannot be
realised as an intersection of the lightcone coordinates of $\Phi(C)$,
for $R$ large. However, it is always possible to make appropriate
changes in the embedding density in order to construct an embedding
$E:C \rightarrow (I,\eta) $: $E$ will not be faithful, but $C$ can be
realised as the intersection of the lightcone coordinates of $E(C)$
and is hence a 2D order.
It would thus appear that the class of 2D orders includes faithful
embeddings into intervals with regions cut out of it, i.e., topologies
different from the disc. On the other hand, it is always possible to
choose an interval spacetime $(I,g)$, with a $g=\Omega(u,v)^2 \eta$
which ``compensates'' for the varying embedding density of $E(C)$ in
$(I,\eta)$, so that $\Phi: C \rightarrow (I,g) $ is a faithful
embedding (see Fig \ref{difftgy} (b))
Relevant to our model, is the resulting statement that a 2D order
which embeds into an interval spacetime with holes can equivalently be
obtained as a discretisation of a conformally flat interval spacetime.

\begin{figure}[ht]
\centering \resizebox{1.5in}{!}{\includegraphics{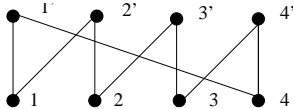}}
\vspace{0.5cm}
\caption{{\small An eight element crown poset, constructed from 
    $A_1=(e_1,e_2,e_3,e_4)$ and $A_2=(e_1',e_2',e_3',e_4')$.}}\label{crown}
\end{figure}

The class of continuum manifolds of typical interest in 2D quantum
gravity are ones with spacelike boundaries representing an initial and
a final time.  From this perspective, the extension of $I$ to $I'$
seems reasonable, since $I'$ has both initial and final spacelike
boundaries.  Now, except spacetimes on the interval topology,  all 2D
spacetimes of finite volume which satisfy this boundary requirement
have non-contractible loops and non-vanishing first Betti numbers
$\beta_1$. $\beta_1$ is also non-vanishing for any spatial slice in
these spacetimes. In \cite{hom} it was shown that causal sets $C$ that
faithfully embed into a globally hyperbolic region of a spacetime
contain sufficient structure to reproduce the spatial continuum
homology with high probability. The construction in \cite{hom} uses
the idea of a thickened antichain from which a nerve simplicial
complex is constructed. In particular, it can be shown that if
$\beta_1 \neq 0$ for this nerve simplex, it implies the existence of a
``crown'' sub-poset in $C$.
A crown poset is defined as follows. Let $C$ have cardinality, $2m$,
$m>2$, and let $A_1=(e_1, e_2, \ldots e_m)$, $A_2=(e_1', e_2', \ldots
e_m') $ be two non-intersecting antichains in $C$, whose elements are
related to each other by  
$e_i \prec e_i',e_{i+1}'$ and $e_i' \succ e_{i-1},e_i$, where we are treating
indices modulo $m$ (see Fig \ref{crown} for an example). We note that: 

\begin{claim}A 2D order cannot contain a crown poset with
  $m>2$. 
\end{claim} 
\bproof Suppose the crown $C_m$ is the intersection of linear orders $L$ and
$M$. Let $e'_i$ be the lowest primed vertex in $L$.  Then $e_i$
 and $e_{i-1}$ both appear below all the primed vertices in $L$.  Now
 suppose wlog that $e_i$ appears above $e_{i-1}$ in $M$.  Then
 $e'_{i+1}$ appears above $e_i$ in $M$, and hence is above $e_{i-1}$
 in $M$ as well as in $L$.  As $e'_{i+1}$ is not above $e_{i-1}$ in
 the crown $C_m$, this contradicts the assertion that $L \cap M =
 C_m$. 
\eproof 

To see this result within the wider context of the theory of poset
dimension, the reader should consult \cite{trotter}.  What one would
like to deduce from this is that 2D orders are exclusively associated
with the interval topology. Of course, if the scale of the topology is
of order the discretisation scale, then this is no longer the
case. However, such continuum structure is considered irrelevant from
the causet perspective, and hence discretisation of such spacetimes is
not pertinent.  Thus, within these limitations we may conclude that if
a 2D order approximates to a 2 dimensional spacetime, then the latter
belongs to the class of conformally flat interval spts. 
It is in this sense that the class of 2D orders distinguishes the
topology of the interval from all the others relevant to 2D quantum
gravity. The set of \textit{all} 2D orders is thus a meaningful causet
discretisation of the class of 2D interval spacetimes.

We are now in a position to write down the causet partition function.
For 2D orders that do have a continuum approximation, our
discretisation gives us a uniform measure coming from the continuum
action (\ref{contaction}). Moreover, since the set of all 2D orders
includes all causets corresponding to 2D intervals, but none
corresponding to other 2D spacetimes, it is natural to extend this
uniform measure on manifoldlike 2D orders to all 2D orders. The
partition function for our model is thus the unweighted sum over the
set $\Omega_{2D}$ of unlabelled 2D orders 
\begin{equation}\label{partition}
\tZ = (\mathrm{phase}) \times  \sum_{\hs} \id.
\end{equation}
The appearance of a uniform weight in the partition function comes
from the triviality of the continuum theory, and at first glance
suggests that any semi-classical regime will be impossible in the
model. In path integral quantum mechanics, for example, the set of all
paths is dominated by non-classical, non-differentiable paths.  The
inclusion of the non-trivial weight $\exp(i\frac{S(\gamma)}{\hbar})$ is
crucial in obtaining the correct classical limit. 

Indeed, as shown in \cite{kr}, a uniform measure over the set of all
$N$ element causets, not just those which are 2D orders, is completely
dominated in the large $N$ limit by the Kleitman-Rothschild
three-level (or three ``moments-of-time'') causets, which are most
non-manifoldlike. It is hoped that a suitable action for causets would
repair this entropic problem and yield the correct continuum
approximation or classical limit.  In our model, however, since the
action is trivial, the partition function is determined solely by
entropic effects. Nevertheless, because our measure vanishes on all
$N$ element causets which are not 2D orders, a meaningful continuum
approximation does indeed emerge from this theory as we will discuss
below.

As labels are the discrete analogues of coordinates they are
considered unphysical in causet theory.  Our interest therefore
lies with isomorphism classes of labelled 2D orders, i.e., with
unlabelled 2D orders. The random variable $U(N)$ on the isomorphism
classes of labelled 2D orders each taken with equal probability 
therefore matches the  normalised partition function for our
model (\ref{partition}). We will also be interested in so-called
labelled {\sl random 2D orders} $P(N)\equiv L \cap M$ which are random
variables defined by choosing $L$ and $M$ randomly and 
independently from the $N!$ linear orderings of
$\{e_1,...,e_N\}$. The study of random $k$-dimensional orders was
initiated in the 1980s by Winkler~\cite{W2}.  The case $k=2$ has
been of particular interest, because of its connection to random
permutations, and to Young tableaux. The typical structure of a random
2-dimensional order is now reasonably well understood.

From the perspective of CSQG, this model of random orders plays a
crucial role.  Indeed, a random order from this model can equivalently
be generated by taking a sequence of $N$ independent random points in
a fixed interval $I$ of 2D Minkowski spacetime, according to the
volume measure \cite{W2}.  This in turn is equivalent to the Poisson
process (or sprinkling) in the interval, conditioned on the number of
points being $N$~\cite{Stoyan}.  The equivalence of the two models can
be seen as follows.

Let $I$ be the interval of 2D Minkowski spacetime between two points
$a$ and $b$, with lightcone coordinates $(u_a,v_a)$ and $(u_b,v_b)$
respectively.  Thus $I$ is the rectangle consisting of all points with
$u$-coordinate in $[u_a,u_b]$ and $v$-coordinate in $[v_a,v_b]$. Now
let $C$ be the causet, with elements $\{e_1, \dots, e_N\}$, obtained
by choosing points $\{\Phi(e_1),...,\Phi(e_N)\}$ independently
uniformly at random from $I$, and taking $C$ to be the induced order:
$e_i \prec e_j$ in $C$ if $\Phi(e_i) < \Phi(e_j)$ in the causal order
on the manifold.  Let $(u_i,v_i)$ denote the coordinates of the
sprinkled element $\Phi(e_i)$.  With probability~1, all the values
$u_i$ and $v_i$ are different.  As described above, $C$ is the
intersection of the linear orders $U$, $V$ obtained from these $u$ and
$v$ values.  Each of the pairs $(u_i,v_i)$ is chosen uniformly over
the rectangle $I$, so the coordinates $u_i$ and $v_i$ are independent
of each other, and of all other choices.  Thus no permutation of the
elements of $C$ can be more likely to occur as the order $U$ than any
other, i.e., the random linear order $U$ is distributed uniformly over
all linear orders of elements of $C$.  The order $V$ is also uniform
over the set of all linear orders, and is independent of $U$.  The
process of taking a sprinkling and deriving a (labelled) causet from
it is therefore equivalent to taking a random causet according to $P(N)$.

This means that a ``typical'' random order from $P(N)$ corresponds (in
the sense of a faithful embedding) to an interval of 2D Minkowski
space of volume $N V_c$.  For a spacetime with non-trivial conformal
factor, while the process of sprinkling is still a random process,
there will in general be correlations in the $u$ and $v$
values. Hence, sprinklings into such spacetimes which differ from flat
spacetime at scales much larger than the cut-off, are not equivalent
to the random 2D orders $P(N)$.

The following result was first proved by El-Zahar and
Sauer~\cite{ElZ-S}, and was stated in this form by
Winkler~\cite{winkler}, who gave an alternative proof and considered
the (more complicated) labelled case as well.
\begin{theorem} \label{winkler}
Let $\Phi$ be an isomorphism-invariant statement about 2D orders which
has a limiting probability   either in $P(N)$ or in $U(N)$. Then a
limiting probability exists in the other case as well and the two
probabilities are equal.
\end{theorem}
The proofs of El-Zahar and Sauer, and of Winkler, also give that the
number of $N$-element 2D orders is $N!/2 (1+o(1))$, and that almost
all of  them have a unique representation, up to isomorphism, as an
intersection  of two linear orders.
Here, ``limiting probability'' refers to the probability in the $N
\rightarrow \infty$ limit. From our discussion above, it then follows
that as $N \rightarrow \infty$, the partition function
(\ref{partition}) is dominated by causets which faithfully embed into
an interval of Minkowski spacetime of volume $V = NV_c$.  This
emergence of manifoldlike causets in an apparently featureless
partition function is surprising, to say the least. Dominance of
a class of configurations in the partition function has a standard
interpretation in quantum theory, which translates in our case to the
statement that {\it 2D Minkowski spacetime is a prediction of our
theory}.

The large $N$ limit taken above can be interpreted as a large volume, if
the discreteness scale $V_c$ is held constant, or a continuum limit
if, instead, the total volume $V$ is held constant. We see in the above
model that, in the continuum approximation, fluctuations die out
altogether,  with flat space dominating. Thus, despite the possibility
of having no classical limit at all in 2D, the continuum approximation
is actually a classical limit.  This is not a feature to be found in
other 2D quantum gravity models \cite{fdavid}.  However, it \textit {is}
something that is desirable in a model of 4D CSQG, since the
discreteness scale is of order the quantum gravity scale. These
results are therefore interesting from this point of view. The size of
quantum fluctuations at given $N$ remains to be calculated and will
require numerical analysis. We leave this for future investigations.

A similar model may be constructed for the cylinder topology $S^1
\times \re$, the other class of 2 dimensional spacetimes with fixed
topology.  Causets on the cylinder can be partly characterised by the
existence of the crown sub-posets described above, resulting in a
non-vanishing first Betti number (which means that they are not 2D
orders). However, we know of no definitive characterisation of such
``cylinder'' posets analogous to the 2D orders discussed above for the
disc topology. It would be of great interest to check if a continuum
spacetime is also emergent for this class of causets. Such a model
would help in a more straightforward comparison with existing 2D
quantum gravity models -- would the radius of the cylinder fluctuate
in the continuum limit as in other models, or would a classical limit
be obtained?

In this model, as in other lower-dimensional models, many of the
problems that exist in the 4D case are avoided, but not always in a
way that immediately suggests answers to the 4D problems.
Nonetheless, there are some lessons from 2D to be learned.  It is an
encouraging and non-trivial fact that, in the set of 2D orders,
sprinklings of flat space naturally dominate. Once the restriction to
2D orders is made, non-manifoldlike causets are no longer entropically
preferred.  In 4D, causets that can be embedded into intervals of
Minkowski space are known as ``4D sphere orders'' \cite{trotter}.
It would be of great interest to know whether any analog of the
El-Zahar/Sauer result holds here: we can define the probability spaces
$U(N)$ and $P(N)$ as in the 2D case, where $P(N)$ now refers to
sprinkling into an interval of 4D Minkowski space, and ask how these
are related.  It is probably too much to expect that every statement
about 4D sphere orders has the same limiting probability in the two
models, but nevertheless it may well be true that a causet drawn from
$U(N)$ typically corresponds to an interval in the manifold, in the
sense considered here.  If so, this bodes well for the entropy problem
in CSQG.

\end{document}